%% file: main.tex
\documentclass[acmtog,nonacm]{acmart}
\acmSubmissionID{1234}

\usepackage{booktabs} 

\usepackage[ruled]{algorithm2e} 

\SetAlFnt{\small}
\SetAlCapFnt{\small}
\SetAlCapNameFnt{\small}
\SetAlCapHSkip{0pt}

\acmJournal{TOG}

\usepackage{xcolor}
\usepackage{lstlinebgrd}

\definecolor{mGreen}{rgb}{0,0.6,0}
\definecolor{mGray}{rgb}{0.5,0.5,0.5}
\definecolor{mPurple}{rgb}{0.58,0,0.82}

\lstdefinestyle{CStyle}{
    commentstyle=\color{mGreen},
    keywordstyle=\color{magenta},
    numberstyle=\tiny\color{mGray},
    stringstyle=\color{mPurple},
    basicstyle=\footnotesize,
    breakatwhitespace=false,         
    breaklines=true,                 
    captionpos=b,                    
    keepspaces=true,                 
    numbers=left,                    
    numbersep=5pt,                  
    showspaces=false,                
    showstringspaces=false,
    showtabs=false,                  
    tabsize=2,
    language=C
}

\DeclareMathOperator{\atan}{atan}

\begin{document}
\title{Composable function systems as a general-purpose rendering framework}

\author{James Schloss}
\orcid{000-0002-3243-8918}
\affiliation{%
  \institution{Massachusetts Institute of Technology}
  \country{USA}}
\email{jars@mit.edu}

\begin{abstract}
Function systems exist as a natural language for the meshless creation and manipulation of complex objects while maintaining minimal memory on the Graphics Processing Unit (GPU) or Central Processing Unit (CPU).
This paper proposes a new method for general-purpose (non-fractal) visualizations and simulations with function systems and introduces Quibble, a metaprogramming framework for composing such systems on the GPU.
We also discuss several core advantages of this method including runtime performance, the creation of topologically non-trivial objects, and interoperability with other graphical algorithms.
Beyond general-purpose imagery and animations, this method can also be used to give artists more control over in-between frames in low-framerate animations, controllably deform point clouds, and metaprogram difficult animation workflows.

\end{abstract}

%
%


%
%


\begin{teaserfigure}
\includegraphics[width=\textwidth]{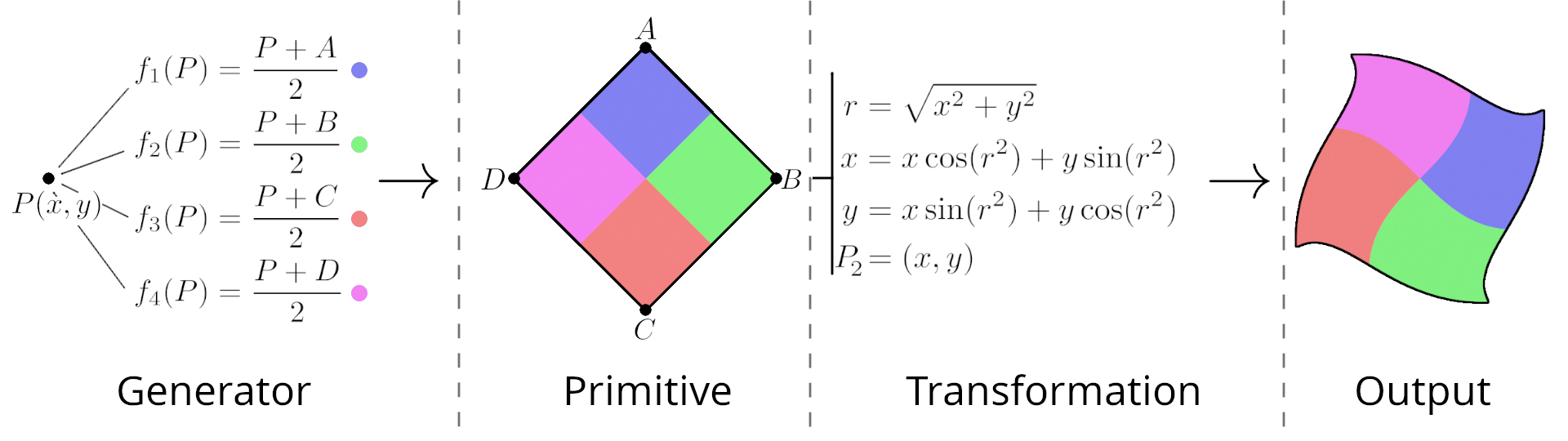}
\caption{A single step in a simple CFS workflow, including the generation of an object primitive with an iterated function system followed by a spatial transformation with human-readable mathematical operation. Note that the initial point ($P$) is distinct from the final output point ($P_2$).}
\label{fig:workflow}
\end{teaserfigure}

\maketitle 

\section{Introduction}

Function systems are powerful tools that have been used to generate mathematical art for decades~\cite{barnsley1985iterated, barnsley1989recurrent, barnsley2003v, elliott2003functional, duff1992interval, keeter2020massively}.
Iterated Function Systems (IFSs), in particular, are likely the most well-studied functional approach for creating and studying self-similar (fractal-like) behaviour~\cite{ghosh2022iterated, diaconis1999iterated, iosifescu2009iterated, bouke2025fractal} and have been use for many exciting applications, such as spatial sampling and fractal image compression~\cite{fisher1994fractal, kominek1995algorithm}.
Solving IFSs is typically done with a random sampling method known as the chaos game, where a single point chooses a single function at random every step \cite{barnsley2011chaos}.
After transformation by the chosen function, the point then colors the appropriate pixel on screen before moving on to the next step.
As the simulation continues, the point drifts towards the final shape, known as the attractor of the system, and after several thousand iterations, the shape should appear on the screen as well.
An example of such a function system is shown in Figure~\ref{fig:workflow} (Generator / Primitive), where the set of four functions ($f_1$, $f_2$, $f_3$, and $f_4$) are used to generate a square primitive for later use.
It is important to note that the square shown in the ``Primitive'' section can be thought of as a point cloud, where each point has been histogrammed or splatted on to some output layer and the colors indicate the function used on each individual point to land on its final position.
Because each function is chosen at random, it is typically difficult for compilers to inline these functions without calling them as function pointers.
Other functional approaches to graphics~\cite{elliott2003functional, duff1992interval} exist with different advantages and drawbacks.
These will be discussed in Section~\ref{sec:related}; however, all approaches are advantageous in that they allow for the creation of complex objects without a large amount of data stored on the Graphics Processing Unit (GPU) as the point cloud to be output to screen does not (necessarily) need to be stored in a global address space such as with \texttt{GL\_POINTS}-based methods often used in photogrammetry and LiDAR\cite{leberl2010point} as well as many other applications\cite{10301359, iglhaut2019structure, du2023advances}.
This means that for many workflows, the only data needed to be stored in GPU global memory is the final output buffer of the image, itself, as well as a few bytes for necessary function arguments.

With the recent resurgence of splatting techniques (in particular Gaussian splatting)\cite{bao20253d} and the relatively high prices of consumer memory, it is relevant to revisit such techniques as a possible methods to generate large scenes stored as point clouds.
Unfortunately, it is difficult to efficiently solve the chaos game on GPU hardware due to a large number of limitations, including warp divergence from random function selection~\cite{howes2007efficient}, the usage of function pointers~\cite{zhang2021judging}, and atomic operations for histogramming the final point location to the output buffer~\cite{shams2007efficient}.
Though there exist several techniques to circumvent these issues \cite{schied2011high}, this work proposes another direction.
Rather than building a single function system to be solved with the chaos game, we instead generalize the creation of different object primitives and allow for users to create a Composable Function System (CFS) framework to specify precisely how all the points in the final cloud are intended to move at later stages in the pipeline.
A single step of an example workflow is  shown in Figure~\ref{fig:workflow}, where an IFS is used for the generation of the starting square before another (notably non-affine) function is used to generate the final object.
The initial step of generating the object primitive and shading it can be generalized to any approach.
It is possible, for example, for a set of objects to be created by photogrammetry and shaded based on lighting in the scene.
The user can then specify how they desire for the points to move on screen without the need to explicitly rig an object.
It is also possible to first create an object primitive with an IFS (for example) and then shade it with another approach, such as signed distance fields~\cite{jones20063d, frisken2006designing}, ray tracing~\cite{glassner1989introduction, slusallek2005introduction}, or other software rendering techniques~\cite{laine2011high}.
In this way, CFSs are highly inter-operable with other, existing methods in the literature.

This paper is organized in the following way.
In Section~\ref{sec:related}, we discuss several related methods to CFSs.
In Section~\ref{sec:intro} we provide a simple example of function composability as well as the software interface (Section~\ref{sec:api}) and a motivating example (Section~\ref{sec:space}).
Finally, we conclude and discuss future work in Section~\ref{sec:conclusion}.

\section{Related work}
\label{sec:related}

IFSs and related approaches have been an active field of research interest for decades, with many methods showcasing their versatility and usability in various aspects.
As such, CFSs build upon and extend several other existing techniques in the literature.
For example, in \cite{kunze2008iterated}, the authors propose extending IFSs to a new technique they call the Iterated Multifunction System, which allows for greater flexibility in the creation of fractal-like objects; however, their approach is limited in terms of allowed mappings.
The fractal flame algorithm \cite{draves2008fractal} is also similar in that it allows for users to inject functions into a pre-existing IFS structure.
Several other approaches limit themselves to affine maps\cite{hutchinson1981fractals, patino2023brief, husain2022fractals}, where no such limitations exist for this work.
There are also methods for rendering that do not necessarily use IFSs, and instead use function systems more generally~\cite{elliott2003functional} or with recursive methods~\cite{keeter2020massively, duff1992interval}.
In contrast to all of of these methods, this work focuses on building a metaprogramming framework to allow users to flexibly mix and match their preferred methods for composing functional units and create general-purpose (non-fractal) artwork, though we focus on IFSs for the generation of object primitives.

For general-purpose rendering, CFSs must also contrast themselves to other methods such as meshes, ray tracing, and raymarching with signed distance fields.
In contrast to traditional mesh-based approaches, this work does not necessitate the storing of any vertex points or textures in memory and is capable of generating $n$-dimensional scenes on-the-fly to be splat on to screen.
Objects and object locations do not necessarily need to be stored in memory either and can instead be directly embedded into the function systems, themselves.
In this way, it is similar to approaches using work-graphs and mesh shaders for realtime rendering~\cite{kuth2024real, kuth2025real}.
If users wish to use mesh-based methods, it is possible to do so by generating triangles and creating the mesh through software rendering approaches~\cite{laine2011high}; however, the software implementation of CFSs written for this work primarily relies on computational kernels (acting as shaders) and lacks certain hardware features like rasterization and raytracing.
It is also possible for users to implement raytracing and raymarching methods as shaders to objects created with CFSs to be manipulated by (for example) non-affine transformations.
This is particularly interesting in that it extends the capabilities of raymarching with signed distance fields by enabling users to create complex smear effects without needing to march through non-Euclidean spaces~\cite{frisken2000adaptively, seyb2019non}.

Finally, as this work introduces Quibble, a GPU metaprogramming framework, it is also important to differentiate it from other, existing graphical and computational programming interfaces.
At the current time, there are few interfaces that allow for flexible and efficient calling of function pointers on the GPU by default~\cite{zhang2021judging}.
This means that implementing generic, user-defined function calls on the GPU can be troublesome for many applications.
Quibble exists as a metaprogramming framework to solve this problem by taking advantage of features from both graphical and computational workflows.
Unlike graphical interfaces, like Vulkan, OpenGL, DirectX, and others, Quibble focuses primarily on the creation of high-quality computational kernels instead of the traditional graphical pipeline with (for example) vertex and fragment shaders.
Unlike computational interfaces, like CUDA~\cite{sanders2010cuda}, KernelAbstractions.jl\cite{churavy2024kernelabstractions}, SyCL\cite{reyes2016sycl}, and kokkos\cite{edwards2014kokkos}, Quibble instead focuses on compiling a large, custom kernel from user-provided functions at runtime.
Though several modern languages, such as Julia, Rust, and C++ provide flexible metaprogramming capabilities, it is difficult to use such tools directly for this application, as we instead need the staged compilation approach commonly found in graphical languages like Vulkan~\cite{sellers2016vulkan} and OpenGL~\cite{neider1993opengl}
This feature is also present in OpenCL~\cite{munshi2009opencl}, which Quibble transpiles to.
For this reason, Quibble is more similar to kernel fusion frameworks~\cite{yadav2025composing, alomairy2024dynamic}.

\section{Brief introduction to composable function systems for rendering}
\label{sec:intro}

Simply put, a function system is a set of functions with defined inputs and outputs.
In the context of rendering, the functions in the system must somehow color pixel coordinates on an output image.
The simplest strategy to accomplish this with traditional software (such as OpenGL or Vulkan) might be to create a fragment shader on some quad spanning the size of the desired image; however, much more complex functions may also exist.
Signed distance fields, for example, are used with raymarching to create complex scenes.
In the case of IFSs, each function ($f_n$) maps some position ($P_{n}(x, y)$) on to some new position ($P_{n+1}(x, y)$).
Most IFSs are solved via the chaos game, which starts with a single point randomly chosen in real space.
Every iteration of the chaos game, a new function is chosen at random and applied to that point.
After, the point is then splat or histogrammed on to screen.
One core advantage of creating objects by IFSs is that there is no need to keep any of the constituent points in global memory on the GPU.

CFS frameworks, on the other hand, do not have any rigorously defined strategy for function solving and are instead intended to allow users to mix and match different rendering methods much more effiently.
An example of this is shown in Figure~\ref{fig:workflow}, where the initial object is generated by an IFS, shaded by a method similar to a fragment shader, and then later manipulated by an arbitrary spatial function.
By providing users the flexibility to choose their preferred rendering and shading method as well as the ability to choose the stage of the rendering pipeline when each function is called, users have unprecedented control over the entire rendering pipeline and can create complex objects without the need to model anything by hand.
This allows for users to create more complex procedural visualizations or effects on non-procedurally created objects and scenes.

As an example, consider the creation of a unit circle centered about the origin with a non-affine IFS:

\begin{equation}
\begin{aligned}
r &= \sqrt{x^2 + y^2} \\
\theta &= \atan(y  / x) \\ \\
\theta_2 &= \pi (r + f_{id}) \\
r_2 &= \sqrt{\frac{\theta}{2\pi}} \\ \\
x_2 &= r^2\cos(\theta_2) \\
y_2 &= r^2\sin(\theta_2) 
\end{aligned}
\label{func:circle}
\end{equation}

\begin{figure}
\begin{centering}
\includegraphics[width=0.4\textwidth]{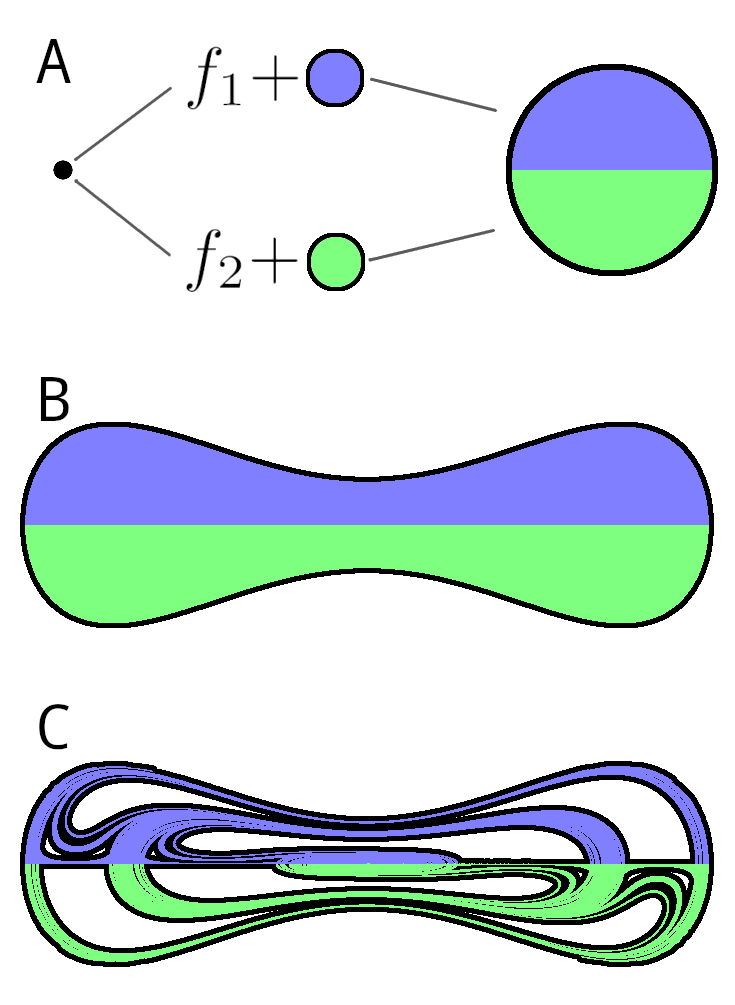}
\end{centering}
\caption{An example of how different methods of function composition can lead to different outcomes. A is the generation of a circle with the equations provided in the text. In B, we use a copy of the point used for generating the primitive before proceeding with additional transformations. In C, we reuse the same point instead.}
\label{fig:circle}
\end{figure}

Here, $x$ and $y$ are the positions of the point, $r$ is the distance from the origin, and $\theta$ is its corresponding angle.
The values $x_2$, $y_2$, $r_2$, and $\theta_2$ represent the values of the next point in the iteration and $f_{id}$ represents either $0$ or $1$ depending on which function is chosen.
This function system can be interpretted as a map from the original circle on to either the top or bottom half of itself.
This is shown in Figure~\ref{fig:circle}(A).

Now imagine that the user would like to transform the circle by stretching it and squeezing in the center with the following transformation:

\begin{equation}
\begin{aligned}
x_2 &= 2x \\
y_2 &= y - \cos(x)
\end{aligned}
\end{equation}

In this case, the user would likely desire the image shown in Figure~\ref{fig:circle}(B).
The simplest way to generate this would be to first generate all the points in the circle, then save those to some array in global memory, and then transform the points separately; however, this might lead to the unnecessary storage of potentially millions of points.
Such a method comes at a high memory cost and should be avoided, but allows for similar staging to traditional mesh-based methods.

\begin{figure*}[ht!]
\input{code\_highlight}

\caption{This shows the compilation of a quibble scribble (left) into an OpenCL kernel (right) and showcases many features of the quibble kernel language. Here, the code is colored based on where it is defined within the quibble scribble. Red is from a \texttt{poem}, blue is from a \texttt{stanza}, and green is from a \texttt{verse}.}
\label{fig:api}
\end{figure*}

\begin{figure*}[ht!]
\includegraphics[width=\textwidth]{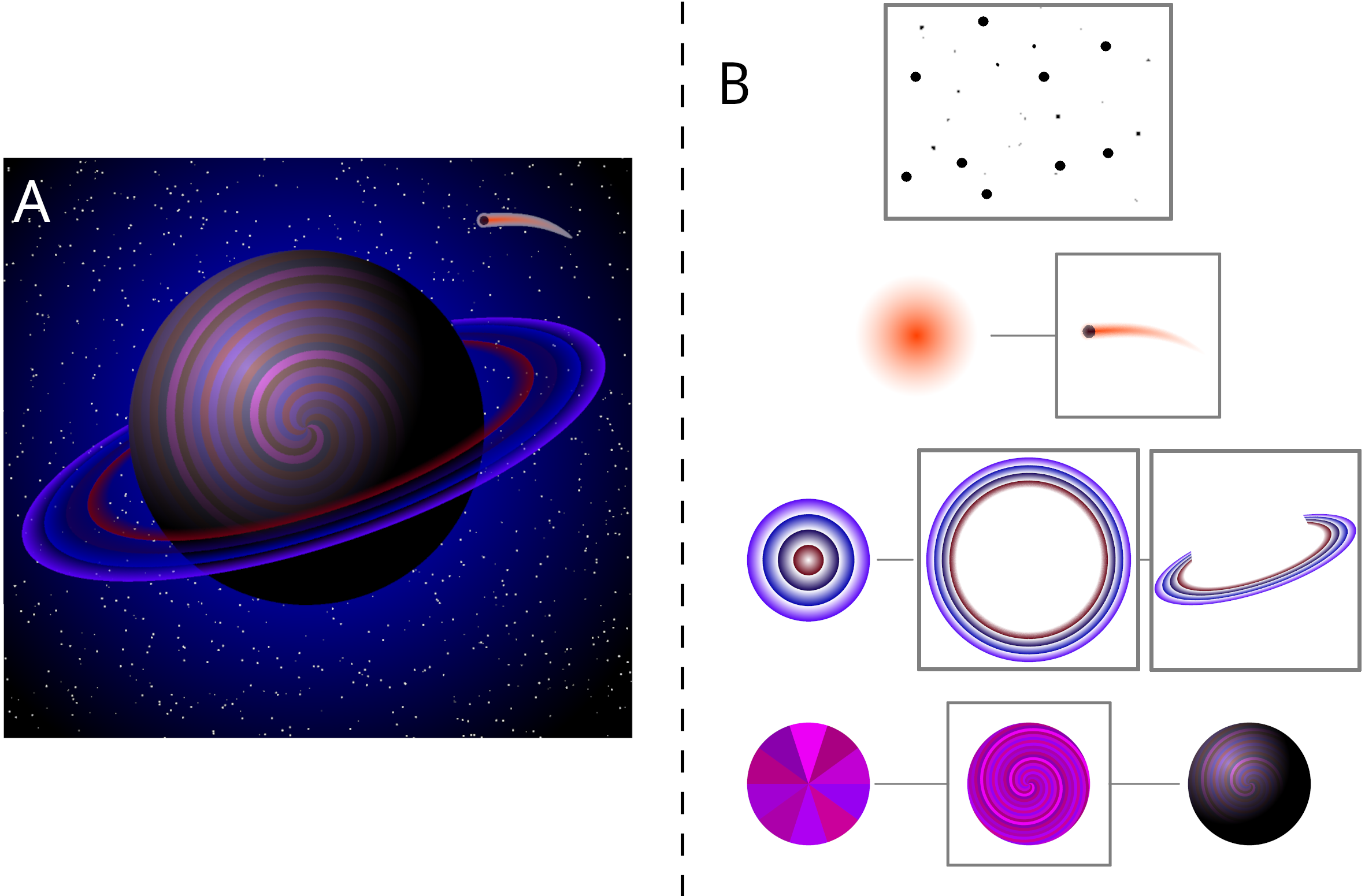}
\caption{A simple render designed with Fable that showcases the capabilities of the current software and interesting features of the algorithm. A is  a render of a planet with rings, stars, and a comet. B shows all the transformations necessary from the generated object primitive to create the final imag e. All objects highlighted in gray boxes are of spatial transformations.}
\label{fig:space}
\end{figure*}

Instead, we should incorporate the transformation function into iteration step.
This can lead to two different results depending on how it is composed with the function set~\ref{func:circle}.
If the user re-uses the point used to generate the circle and outputs that to screen, it will result in Figure~\ref{fig:circle}(C); however, if the user creates a copy of the point and instead transforms that, then they can still reuse their original point for generating the next point in the circle (or other arbitrary primitive) while also allowing for arbitrary transformations before splatting to screen and create the desired image in Figure~\ref{fig:circle}(B).
A more complex example will be discussed in Section~\ref{sec:space} in the generation of Figure~\ref{fig:space}; however, it should be clear that additional flexibility in the rendering pipeline allows for more complex imagery to be created.

\subsection{API overview and motivating examples}
\label{sec:api}

Quibble is intentionally simple and introduces a new set of kernel semantics called \textit{scribbles}, which are interoperable with OpenCL kernel syntax.
This means that scribbles are written in simple C with a few user-directed guidelines to more easily compose function systems together.
This section is not intended to provide an exhaustive API for quibble, but is instead meant to showcase interesting and useful features that will later be expanded upon.
Namely, Quibble introduces:

\begin{description}
\item[\texttt{verse}s] these are function-like objects that allow for users to express operations without compiling them immediately. \texttt{verse}s can be called from within other scribble constructs with the \texttt{@VCALL} macro.
\item[\texttt{stanza}s] these are similar to \texttt{verse}s, but allow for users to use an additional \texttt{\_\_split\_stanza()} function to split the block and inject additional operations in to the \texttt{stanza}. These additional operations may be \texttt{verse}s or any other syntactically correct C code. \texttt{stanza}s can be called within \texttt{poem}s with the \texttt{@SCALL} macro.
\item[\texttt{poem}s] these are directives that compose \texttt{verse}s and \texttt{stanza}s together into OpenCL kernels.
\item[\texttt{@include} directives] to easily compose different scribbles together by combining them in \texttt{.qbl} files, similar to those used in languages like Python and Julia.
\end{description}

Figure~\ref{fig:api} shows an example of many of these scribble constructs and how they transpile into an OpenCL kernel.
Both \texttt{verse}s and \texttt{stanza}s allow for key-word arguments and may use variables defined from within the \texttt{poem} context from which they are called.
Aside from these modifications, \texttt{verse}s are similar in functionality to inlined functions.
All scribble features can be compiled at the user's request in C by either calling on external \texttt{.qbl} files or in a single-source style by using the \texttt{QBINLINE} macro.
By delaying compilation, users can inject complex transformations (with either functions or \texttt{verse}s) into \texttt{stanza}s to compose function systems together and create complex and dynamic scenes.
This functionality is essential to creating a CFS system.

Quibble provides several data structures to work with point clouds, colors, and images, as well as their corresponding data structures in the C api.
There are also pre-made scribbles for standard operations users might request, such as file input and output, random number generation, and the creation of several object primitives like triangles, circles, and squares with IFSs.
From C, Quibble provides a more flexible user-interface than OpenCL with helper functions to (for example) more easily set arguments, transfer data to and from the device, and run programs.
Full examples of the C code and corresponding quibble scribbles for generating shaders, fractals, and smears are all available in the documentation~\cite{quibbledocs}.
Now is a good time to introduce a motivating example as to why these language constructs are helpful.

\subsection{Multiple objects from a single primitive}
\label{sec:space}

Non-photorealistic rendering with mathematical expressions is a popular area of commercial and hobby interest, but software frameworks for such activities are often limited in their design and usability.
Vector-based libraries (such as manim~\cite{zhang2025manim, manimdocs}) are often difficult to parallelize and are non performant for real-time applications.
Shader-based libraries (such as ShaderToy~\cite{quilez2017shadertoy}) require a deeper understanding of the graphical pipeline, which is not necessary for visualizations made through mathematical expressions.
In addition, the mathematics necessary for shader-based approaches is often unintuitive, as in the case of signed distance fields.
CFSs are in a unique position in that the transformations for each object primitive can be both easily parallelized and simply understood based on the user's preference.
With CFSs, it is possible to create a point cloud representing single object (such as a square, triangle, or circle) on-the-fly and re-use the points from this object to create many other ojects in a scene without allocations in global memory beyond the final output image.

One example of this is shown in Figure~\ref{fig:space}.
The primitive was constructed with an IFS chaos game for the circle, shown in Figure~\ref{fig:circle}(A).
After moving a point to the next position in the chaos game, but before writing to screen, we repositioned that point draw the stars, rings, planet, and comet (all shown in Figure~\ref{fig:space}(B)).

Each of the objects in the final image showcase an interesting transformation that can be challenging for traditional methods.
The stars show duplication, where all star locations can be generated on-the-fly with a simple PRNG scheme and do not require explicit allocation on individual threads.
The comet shows a somewhat non-trivial smear.
The planet shows how users can use spatial manipulations to create complex textures as well, efectively mixing the capabilities of both fragment and vertex shaders.
Finally, the rings show the creation topologically non-trivial object, which can be difficult to do with mesh or geometry shaders.

The approximate runtimes for generating Figure~\ref{fig:space} on consumer hardware (AMD 6700 XT) are approximately 0.0215s (average of 1024 runs) or ~45 frames per second, with the wall time for the generation of a single image at approximately 0.745s (average of 1024 runs).
Here, we note that this is using a naive chaos game implementation and the run time has the potential to be sped up significantly with more advanced methods.
Currently, the compilation time comprises the largest portion of the total wall time for the generation of a single image.
Though compilation is slow, it is still possible to consider on-the-fly compilation for certain situations in live environments.

Overall, this example shows how it is possible to create complex scenes from relatively simple object primitives while allocating minimal memory on each thread.
In this case, there were less than ten floating-point values stored per thread and no additional data allocated on global memory except for what was required of the final output image ($1920\times1080$ \texttt{RGBA8888} values or roughly 16 MiB).


%
%
%
%

\section{Conclusion}
\label{sec:conclusion}

In this paper, we have introduced a new technique for the creation of general-purpose animations with composable function systems (CFSs) alongside Quibble, a GPU-accelerated metaprogramming framework for this purpose.
We have provided some motivating examples (Figure~\ref{fig:circle} and figure~\ref{fig:space}) that highlight the utility of this work.
This technique has the ability to produce interesting, non-affine, and even topologically non-trivial transformations on any object defined in a scene and is therefore particularly useful for smear frame and in-between generation for rendering methods where such techniques are otherwise difficult.
In the future, it will be interesting to combine this work with other point-based rendering methods used in (for example) photogrammetry to store $n$-dimensional point clouds as function systems for later manipulation in software.

Even so, there several core limitations of this work.
For example, even though the generation of stand-alone images (without accounting for file output) is fast enough to consider this method as a promising candidate for real-time graphical work, Quibble currently transpiles to OpenCL which is inconvenient for game development or other real-time, interactive experiences.
In addition, rendering environments with density estimation, such as with fractal flame generation and caustics, still require atomic operations, which can also limit performance.
Similarly, this method often requires users to over-populate the point cloud representing the object primitive to account for transformations of that object with additional functions.
Because the function systems are all well-defined, it should be possible to automatically generate the appropriate initial distribution of points in an object primitive to account for all possible transformations with tools similar to autodifferentiation~\cite{moses2020instead}.
All of these are limitations on the overall performance of the method and are topics of further investigation.

\begin{acks}

This work will be submitted to the \textit{NEDO Challenge, Quantum Computing ``Solve Social Issues !''} for the C-9 category \textit{Provision of New Rendering Environment} and has not received direct financial funding.
The work has also been developed alongside the LeiosLabs community of passionate programmers on twitch, youtube, github, etc.
I would like to thank Dr. Valentin Churavy for continued support and discussion regarding Julia, LLVM, Autodifferentiation and more.

\end{acks}

\bibliographystyle{ACM-Reference-Format}
\bibliography{literature}

\end{document}

%% file: code_highlight.tex
\begin{minipage}{0.48\textwidth}
\begin{lstlisting}[style=CStyle, linebackgroundcolor={%
    \ifnum\value{lstnumber}<5
            \color{green!25}
    \fi
    \ifnum\value{lstnumber} > 5 \ifnum\value{lstnumber} < 16
            \color{blue!25}
    \fi\fi
    \ifnum\value{lstnumber} > 16 \ifnum\value{lstnumber} < 31
            \color{red!25}
    \fi\fi
    }
    ]
__verse mult_by(__global float *verse_a |
                float factor = 8;){
    verse_a[_idx] *= factor;
}

__stanza stanza_check(__global float *stanza_a,
                      __global float *stanza_b,
                      __global float *stanza_c |
                      float r = 1.0;){
    if (_idx < array_size){
        stanza_c[_idx] = stanza_a[_idx] + stanza_b[_idx];
        __split_stanza();
        stanza_c[_idx] += 1;
    }
}

__poem poem_check(__global float *a,
                  __global float *b,
                  __global float *c,
                  int stanza_num, 
                  int array_size){
    if (stanza_num == 0){
        @SCALL stanza_check(a, b, c | r = 5;){
            @VCALL mult_by(c | factor = 2;);
        }
        if (_idx < array_size){
            @VCALL mult_by(c | factor = 0.5;);
        }
    }
}
\end{lstlisting}
\end{minipage}
\begin{minipage}{0.03\textwidth}
\hspace{0.03\textwidth}
\end{minipage}
\begin{minipage}{0.48\textwidth}
\begin{lstlisting}[style=CStyle, linebackgroundcolor={%
    \ifnum\value{lstnumber}<6
            \color{red!25}
    \fi
    \ifnum\value{lstnumber}=8
            \color{red!25}
    \fi
    \ifnum\value{lstnumber}=30
            \color{red!25}
    \fi
    \ifnum\value{lstnumber}=37
            \color{red!25}
    \fi
    \ifnum\value{lstnumber} > 9 \ifnum\value{lstnumber} < 19
            \color{blue!25}
    \fi\fi
    \ifnum\value{lstnumber}=26
            \color{blue!25}
    \fi
    \ifnum\value{lstnumber}=27
            \color{blue!25}
    \fi
    \ifnum\value{lstnumber} > 19 \ifnum\value{lstnumber} < 24
            \color{green!25}
    \fi\fi
    \ifnum\value{lstnumber} > 31 \ifnum\value{lstnumber} < 36
            \color{green!25}
    \fi\fi
    }
    ]
__kernel void poem_check(__global float * a,
                         __global float * b,
                         __global float * c,
                         int stanza_num,
                         int array_size){
    int _idx = get_global_id(0);

    if (stanza_num == 0){
        {
            __global float *stanza_a = a;
            __global float *stanza_b = b;
            __global float *stanza_c = c;
            float r = 5;

            if (_idx < array_size){
                stanza_c[_idx] =
                    stanza_a[_idx] + stanza_b[_idx];
        
                {
                    __global float *verse_a = c;
                    float factor = 2;

                    verse_a[_idx] *= factor;
                }

                stanza_c[_idx] += 1;
            }
        }

        if (_idx < array_size){
            {
                __global float *verse_a = c;
                float factor = 0.5;

                verse_a[_idx] *= factor;
            }
        }
    }
}
\end{lstlisting}
\end{minipage}